\documentclass[reprint,superscriptaddress,amsmath,amssymb,aps,prl]{revtex4-2}
\usepackage{graphicx}
\usepackage{dcolumn}
\usepackage{bm}
\usepackage[colorlinks,allcolors=blue]{hyperref}
\usepackage{color}
\usepackage{gensymb}
\usepackage{natbib}
\usepackage{txfonts}
\usepackage[final]{changes}
\usepackage{changes}

\begin{document}
\preprint{}

\title{Reversible fully spin polarization in strain-engineered two-dimensional fully compensated magnets}

\author{Xiuli Zhang}
\affiliation{Laboratory for Quantum Design of Functional Materials, and School of Physics and Electronic Engineering, Jiangsu Normal University, Xuzhou 221116, China}

\author{Peng Jiang}
\email{pjiang@jsnu.edu.cn;pjiang93@mail.ustc.edu.cn}
\affiliation{Laboratory for Quantum Design of Functional Materials, and School of Physics and Electronic Engineering, Jiangsu Normal University, Xuzhou 221116, China}

\author{Yurui Ma}
\affiliation{Laboratory for Quantum Design of Functional Materials, and School of Physics and Electronic Engineering, Jiangsu Normal University, Xuzhou 221116, China}

\author{Xiaodong Zhou}
\email{zhouxiaodong@tiangong.edu.cn}
\affiliation{School of Physical Science and Technology, Tiangong University, Tianjin, China}

\author{Linlin Liu}
\affiliation{Laboratory for Quantum Design of Functional Materials, and School of Physics and Electronic Engineering, Jiangsu Normal University, Xuzhou 221116, China}

\author{Hong-Mei Huang}
\affiliation{Laboratory for Quantum Design of Functional Materials, and School of Physics and Electronic Engineering, Jiangsu Normal University, Xuzhou 221116, China}

\author{San-Dong Guo}
\affiliation{
School of Electronic Engineering, Xi'an University of Posts and Telecommunications, Xi'an 710121, China}

\author{Tengfei Cao}
\affiliation{Department of Materials Science and Engineering,
Northwestern Polytechnical University, Xi‘an 710072, China}

\author{Yan-Ling Li}
\email{ylli@jsnu.edu.cn}
\affiliation{Laboratory for Quantum Design of Functional Materials, and School of Physics and Electronic Engineering, Jiangsu Normal University, Xuzhou 221116, China}

\begin{abstract}
Achieving controllable spin polarization and its reversal simultaneously 
in symmetry-compensated magnets is precluded by crystalline symmetries.
Here we demonstrate, using symmetry analysis and a minimal tight-binding model, that uniaxial strain removes 
these constraints by inducing inequivalence between 
magnetic sublattices in two-dimensional (2D) system, driving an altermagnetic (AM) state into 
a fully compensated ferrimagnetic (fFIM) state
and enabling fully spin polarization. 
Furthermore, strain along orthogonal directions gives rise to two energetically degenerate fFIM states with 
opposite spin polarization, enabling reversible spin switching.
More importantly, the two symmetry-related fFIM states can be regarded as distinct ferroelastic variants, 
suggesting that this model or mechanism can be extended to ferroelastic fFIM systems. 
The generality of this mechanism is confirmed by combining spin-group analysis, first-principles calculations, 
and Boltzmann transport theory in representative candidates, including AM Mn$_2$SeO 
and ferroelastic fFIM V$_2$SO. 
Our results reveal a universal symmetry-driven framework for strain-controlled and -reversible 
fully spin-polarized transport and identify strain-engineered AM and ferroelastic fFIM systems 
as a promising platform for  volatile and nonvolatile spintronic applications.
\end{abstract}

\maketitle
{\it Introduction}-
Efficient generation and manipulation of spin polarized current underpin the operation of modern spintronic devices
\cite{RevModPhys.76.323,2007The,https://doi.org/10.1002/adma.202505779,shao2021spin,PhysRevB.110.174429}.
This functionality traditionally relies on ferromagnets, where intrinsic spin polarization
originates from exchange interactions that lifts spin degeneracy as a consequence of 
broken time-reversal ($\mathcal{T}$) symmetry.
However, the accompanying net magnetization inevitably introduces stray magnetic fields
and restricts operation speeds \cite{2007The,https://doi.org/10.1002/adma.202505779}.
Conventional antiferromagnets, on the other hand, provide a magnetically compensated alternative without stray fields 
and with faster switching dynamics \cite{RevModPhys.90.015005}. Unfortunately, their electronic structures remain spin-degenerate 
due to the combinations of $\mathcal{T}$ and spatial symmetries including 
inversion ($\mathcal{P}$) or translation ($\textbf{t}$), hindering efficient spin-polarized transport
\cite{PhysRevB.102.245417,PhysRevB.106.245423}.
Beyond the conventional ferromagnetic (FM) and antiferromagnetic (AFM) paradigms, 
altermagnetism has recently been identified as a distinct magnetic phase \cite{PhysRevX.12.031042,
hayami2019momentum,
PhysRevB.102.014422,ma2021multifunctional,
PhysRevB.109.115102,PhysRevLett.130.216701,reimers2024direct,zhang2025giant,PhysRevLett.134.106801,doi:10.1021/acs.nanolett.5c02121}. 
Despite possessing fully compensated collinear magnetic moments, 
altermagnets exhibit intrinsic spin splitting in the momentum space, 
thereby enabling intrinsic spin splitting states, a feature typically associated with ferromagnets 
but realized here without net magnetization. Such behavior originates from the exchange interactions between
spin sublattices connected by crystal symmetry ($C$) operations that are not $\mathcal{P}$ and $\textbf{t}$ \cite{PhysRevX.12.031042,ma2021multifunctional}.

From an application perspective, fully spin polarization and its reversible control are highly desirable for robust and energy-efficient spintronic devices, particularly in two-dimensional (2D) altermagnetic (AM) materials 
compatible with nanoscale integration \cite{RevModPhys.76.323,2007The,https://doi.org/10.1002/adma.202505779}. 
It is a fact that realizing these functionalities simultaneously 
remains highly challenging in AM systems.
On the one hand, $C$-paired spin-momentum locking inherent to altermagnets impose a 
constraint on the electronic structure, 
whereby spin-up and spin-down states are related by $C$-paired symmetry and remain energetically 
degenerate \cite{ma2021multifunctional,song2025altermagnets}, thereby prohibiting fully spin polarization.
Furthermore, although altermagnets exhibit large spin polarization, 
its anisotropy imposes a dependence on the direction of the external electric field \cite{zhang2025giant,w52v-blqm}.
On the other hand, controllable reversal of spin polarization in magnetically compensated AFM or 
AM systems is generally tied to the reorientation of the Néel order. Practically, it is difficult to
realize because vanishing net magnetization suppresses magnetic field coupling, 
making the reversal of the magnetic moment much harder than in ferromagnets \cite{fkyr-z5b8,doi:10.1126/sciadv.adn0479}.
Although spin–orbit torques have enabled reorientation of the Néel vector, 
they typically induce only partial in-plane rotations (such as 90\degree~or 120\degree) 
\cite{bodnar2018writing,PhysRevLett.124.027202},
which require complex multi-terminal geometries and thus complicate device integration.
Under these constraints, several symmetry-breaking strategies, such as electric fields \cite{PhysRevLett.133.056401,qnwp-8sx5}, 
and spin-ordering engineering \cite{vzh9-gldl}, have sequentially been proposed to break $C$-paired symmetry constraints in altermagnets, induce a transition from the 
AM to a fully compensated ferrimagnetic (fFIM) state, 
and consequently realize high spin polarization.
Meanwhile, recent studies have shown that uniaxial strain engineering, via magnetoelastic coupling between 
ferroelastic distortions and magnetic order \cite{jql2-b4c4,ftmr-bh9k,m33v-xwn3,t774-62jc,peng2026ferroelastic}, enables efficient generation and 
reversible switching of spin polarization in 2D AM systems. 
However, the underlying mechanisms and general conditions to the generation and switching of 
spin polarization by the uniaxial strain in AM or fFIM systems remain elusive.

\begin{figure*}[htp]
	\centering
	\includegraphics[width=0.85 \linewidth]{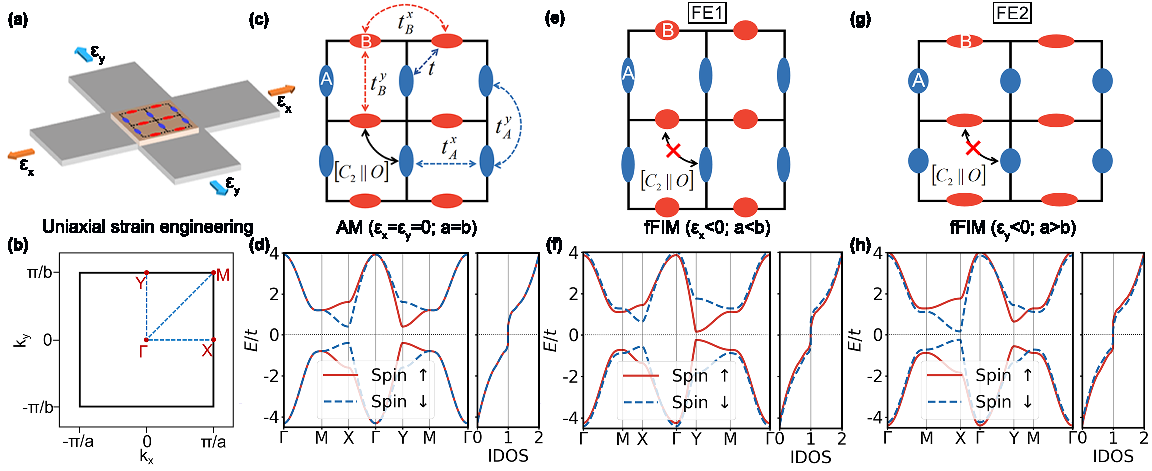}
	\caption{(a) The illustration of uniaxial strain engineering 
	in 2D AM and fFIM systems. (b) First Brillouin zone (BZ)
of 2D square and rectangular lattices. (c) Schematic of a 2D two-sublattice AM model and
(d) its corresponding spin-polarized band structure and IDOS with \( \Delta = 0 \) and \( t_A^x = t_B^y = 0.1t \), and \( t_A^y = t_B^x = -0.2t \), where sublattices $A$ and $B$ are located 
at (0.0, 0.5) and (0.5, 0.0), respectively, 
carrying opposite magnetic moments and interconnected 
by [$C_2 \Vert \mathcal{O}$] ($\mathcal{O}$= $C_{4z}$ or $M_{xy}$) symmetry, in the absence of $\textbf{t}$ and $\mathcal{P}$ symmetries.
(e, g) Schematics of a 2D AM model under compressive strain along the $x$ and $y$ directions, also corresponding to ferroelastic fFIM states FE1 ($a<b$) and FE2 ($a>b$), respectively, 
where sublattices $A$ and $B$ are no longer connected by [$C_2 \Vert \mathcal{O}$].
(f, h) Corresponding spin-polarized band structures and IDOS for FE1 and FE2 configurations 
with (f) $\Delta = -0.2t$; $t_A^x = 0.12t$; $t_A^y = -0.16t$; $t_B^x = -0.24t$; $t_B^y = 0.08t$
and (h) $\Delta = 0.2t$; $t_A^x = 0.08t$; $t_A^y = -0.24t$; $t_B^x = -0.16t$; $t_B^y = 0.12t$ ,
respectively.
	}
	\label{fig1}
\end{figure*}

In this Letter, we develop a symmetry-based minimal model to elucidate strain-induced 
AM–fFIM transitions and spin polarization control. It is demonstrated 
that uniaxial strain lifts energy-resolved spin degeneracy in 2D AM systems and 
stabilizes fully spin-polarized states, with orthogonal strain directions 
providing a symmetry-related route to reverse the spin polarization.
Beyond this, symmetry-related strain configurations map onto two ferroelastic 
fFIM variants with opposite spin polarization, enabling deterministic 
switching between $\pm$100\% spin-polarized transport. 
This mechanism can be well confirmed in 2D AM Mn$_2$SeO and ferroelastic fFIM V$_2$SO 
via first-principles calculations combined with transport analysis.

{\it Model}-To gain microscopic insight into strain-induced full spin polarization and 
controllable reversal, we construct a generic tight-binding model 
based on 2D square and rectangular lattices, which captures AM and fFIM phases. 
The Hamiltonian can be written as \cite{PhysRevLett.134.116703}:
\begin{equation*}
        \begin{aligned}
                H = t &\sum_{i, d_j} \left( c_{A, i}^{\dagger} c_{B, i + d_j} + \text{h.c.} \right) \\
                &+ \sum_{\alpha, i} \left( t_{\alpha}^{x} c_{\alpha, i}^{\dagger} c_{\alpha, i + a_x} + t_{\alpha}^{y} c_{\alpha, i}^{\dagger} c_{\alpha, i + b_y} + \text{h.c.} \right) \\
                &+ \Delta \sum_{\alpha, i} (-1)^{\alpha} c_{\alpha, i}^{\dagger} c_{\alpha, i} + M \sum_{\alpha, i} (-1)^{\alpha} c_{\alpha, i}^{\dagger} c_{\alpha, i} s_z,
        \end{aligned}
\end{equation*}
where $\alpha$ is the sublattice index $A$ or $B$, and
$t$ represents the diagonal nearest-neighbor (NN) hopping parameter
between $A$ and $B$ sites. The next-nearest-neighbor (NNN) hoppings 
between same sublattices are given by $t_{A/B}^{x/y}$.
$d_i$ denotes the NNN vector index, while $a_{x}$/$b_{y}$ represents the lattice 
vector index along the $x$/$y$ direction.
$\Delta$, $s_z$, and $M$ denote staggered potential, spin operator, and magnetization, respectively. 

Figure~\ref{fig1}(a) illustrates uniaxial strain engineering, realized through tensile or 
compressive substrate deformation, enabling control the physical properties of the studied system.
For the pristine or unstrained AM system [see Fig.~\ref{fig1}(c)], 
it satisfies $\Delta = 0$, $t_{A}^{x} = t_{B}^{y}$, $t_{A}^{y} = t_{B}^{x}$, 
and $t_{A/B}^{x} \ne t_{A/B}^{y}$, 
with [$C_2 \Vert \mathcal{P}$] symmetry broken 
and [$C_2 \Vert \mathcal{O}$] symmetry preserved
between sublattices $A$ and $B$.
Therefore, the pristine AM system exhibits anisotropic $d$-wave spin splitting 
in momentum space, while [$C_2 \Vert M_{xy}$] symmetry enforces 
spin degeneracy along the $\Gamma$–M direction, as present in 
Figs.~\ref{fig1}(b) and ~\ref{fig1}(d).
From spin-polarized integrated density of states (IDOS) and band structure, 
one can find that at any given energy the populations of spin-up 
and spin-down states are exactly balanced, 
indicating that there is no net spin polarization in the energy-resolved space, 
akin to conventional AFM systems \cite{jiang2021two,Liu_2024}.
Note that, lifting the spin degeneracy at a given energy is a key requirement for realizing functional spintronic devices \cite{PhysRevB.110.174429}. Although AM materials 
can exhibit large spin polarization, it is highly dependent on the direction of 
the applied electric field \cite{zhang2025giant,w52v-blqm}.
To overcome this limitation, we employ uniaxial strain to break the prime symmetry 
associating sublattices $A$ and $B$, generating a finite staggered potential 
$\Delta$ without introducing extrinsic magnetism. 
This approach can break the symmetry constraint without introducing extrinsic 
magnetic species, preserving the fully compensated magnetic background 
and providing a prerequisite for realizing fully spin polarization.

As shown in Fig.~\ref{fig1}(e) and ~\ref{fig1}(g), we consider two configurations in the 
AM system under uniaxial compressive strain applied along the $x$ and $y$ directions 
($\varepsilon_x < 0$ and $\varepsilon_y < 0$), 
denoted as FE1 and FE2, respectively.
Clearly, the uniaxial strain can change both 
lattice structure and spin-density distribution, which inevitably breaks the pristine $[C_2 \Vert \mathcal{O}]$ symmetry. 
This symmetry breaking removes the protection of the AM phase and drives the emergence 
of a fFIM state, as evidenced in Figs.~\ref{fig1}(f) and~\ref{fig1}(h).
For the FE1 configuration shown in Fig.~\ref{fig1}(f), a clear spin splitting appears 
between spin-up and 
spin-down bands, with the spin-up state dominating near the Fermi level ($E_\mathrm{F}$), 
leading to a fully spin-polarized energy 
window along M-Y-$\Gamma$ paths near $E_\mathrm{F}$ analogous to 
a half-semiconductor \cite{li2016first,PhysRevB.106.245423}. 
IDOS analysis shows that the compressive $\varepsilon_x$ induces a spin-dependent 
redistribution of electronic occupation. Nevertheless, the integrated occupation 
at the $E_\mathrm{F}$ remains spin balanced, yielding zero net magnetization
[see Fig.~\ref{fig1}(f)]. 
This coexistence of spin-dependent occupation and magnetic compensation is a hallmark of 
fFIM feature \cite{qnwp-8sx5,PhysRevLett.134.116703,guo2026sliding}. 
For the FE2 configuration, a compressive strain applied along the $y$ 
direction still induces the AM-to-fFIM transition; 
however, the spin polarization is reversed relative to the FE1 case, 
resulting in spin-down–dominated full polarization along M-X-$\Gamma$ paths 
near the $E_\mathrm{F}$ [see Fig.~\ref{fig1}(h)].
Based on the contrasting behaviors of FE1 and FE2 configurations, 
uniaxial strain engineering enables 
both full spin polarization and its reversible switching through strain-direction control. 
Since these two configurations correspond to distinct ferroelastic phases of the fFIM phase, 
ferroelastic switching provides an effective handle to tune spin splitting \cite{https://doi.org/10.1002/adfm.202525986}. 
This mechanism can be perfectly demonstrated in realistic materials 
using first-principles calculations combined with Boltzmann transport theory, and
more calculation details can be found in the Supplemental Material (SM) \cite{supp}.

\begin{figure}[htbp]
        \centering
        \includegraphics[width=0.95 \linewidth]{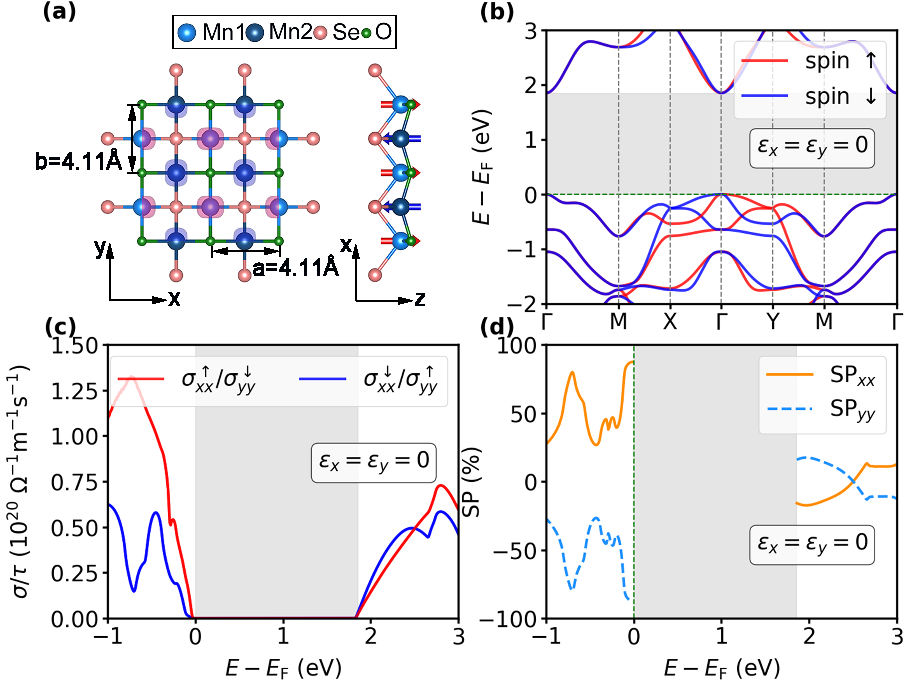}
        \caption{(a) Atomic structure, spin density, and (b) spin-polarized band structure 
        of the 2D ML-Mn$_2$SeO. 
	(c) Spin-resolved longitudinal conductivities and (d) spin polarization ($\mathrm{SP}$) as a function of the energy in ML-Mn$_2$SeO along
	 $x$ and $y$ directions. The light gray shaded area represents the band gap region. $E_\mathrm{F}$ 
	 is set to the energy of the VBM.
  }\label{fig2}
\end{figure}

{\it Material candidates.}-\textbf{AM material Mn$_2$SeO}. 
The monolayer Mn$_2$SeO (ML-Mn$_2$SeO) adopts a tetragonal (square) lattice and 
crystallizes in the $P4mm$ space group (No.~99) with $C_{4v}$ point-group symmetry. 
It can be obtained by removing one of the Se atomic layers 
from the monolayer Mn$_2$Se$_2$O \cite{PhysRevB.110.174429}.
The loss of out-of-plane mirror symmetry reduces the structural symmetry, while maintaining
fourfold rotation $C_{4z}$, twofold rotations $C_{2}$, 
and mirror operations such as $M_{xy}$ and $M_{x\bar{y}}$.
The optimized lattice constants are $a=b$ = 4.11~{\AA}.
Our calculations show that the AM ground state of ML-Mn$_2$SeO is energetically favored [see Fig.~\ref{fig2}(a) and \textcolor{blue}{Fig.~S1} \cite{supp}]. 
The calculated phonon spectra and {\it ab initio} molecular dynamics (AIMD) simulations 
demonstrate both the dynamical stability and thermal robustness of ML-Mn$_2$SeO (see \textcolor{blue}{Fig.~S2} \cite{supp}).
The magnetic moments are primarily localized on the Mn atoms, with a high-spin state of 
approximately 4.48~$\mu_B$, 
and form a collinear order with the Néel vector oriented along the $z$ direction (see \textcolor{blue}{Fig.~S3} \cite{supp}).
With a Néel transition temperature as high as 316~K (see \textcolor{blue}{Fig.~S4} \cite{supp}), 
it is promising for room-temperature spintronic applications.
Figure~\ref{fig2}(b) shows the spin-polarized band structure of ML-Mn$_2$SeO, 
exhibiting a direct-gap semiconducting characteristic with a gap of $\sim 1.85$ eV,
where both valence band maximum (VBM) and conduction band minimum (CBM) are located at $\Gamma$ point.
As expected, the energy bands remain spin degenerate along $\Gamma$–M path, 
while a clear momentum-dependent spin splitting develops along the M–X–$\Gamma$–Y–M path, 
all protected by composite [$C_2 \Vert C_{4z}$] and [$C_2 \Vert M_{xy}$] symmetries.

Next, to further elucidate the spin-polarized transport properties of ML-Mn$_2$SeO, 
we calculate spin-resolved longitudinal conductivities $\sigma_{xx}^{s}$ and 
$\sigma_{yy}^{s}$ ($s=\uparrow$ or $\downarrow$) along the 
principal $x$ and $y$ directions, as shown in Fig.~\ref{fig2}(c).
For a given energy, we consistently find $\sigma_{xx}^{\uparrow} = \sigma_{yy}^{\downarrow}$ 
and $\sigma_{yy}^{\uparrow} = \sigma_{xx}^{\downarrow}$, which is 
enforced by [$C_2 \Vert C_{4z}$] symmetry.
To quantify the spin polarization, its degree is quantified by
$\mathrm{SP}_\mathbf{nn} = \frac{\sigma_\mathbf{nn}^{\uparrow} - \sigma_\mathbf{nn}^{\downarrow}}{\sigma_\mathbf{nn}^{\uparrow} + \sigma_\mathbf{nn}^{\downarrow}}$,
where $\mathbf{n}=(\cos\varphi, \sin\varphi)$ defines the in-plane transport direction, 
and $\varphi$ is the azimuthal angle measured from the $x$ axis.
Thus, a positive (negative) value of $\mathrm{SP}$ indicates 
that transport is dominated by spin-up (spin-down) carriers.
As shown in Figs.~\ref{fig2}(c) and~\ref{fig2}(d), the electric field along $x$ direction 
yields spin-up dominated transport below the $E_\mathrm{F}$ and opposite behavior above it, 
giving rise to positive and negative SP, respectively. For transport along $y$ direction, 
both the dominant spin channel and the sign of $\mathrm{SP}$ are 
reversed due to the [$C_2 \Vert C_{4z}$] symmetry.
Notably, the spin polarization (SP) at the VBM reaches ~90\%, 
but only for electric fields applied along the principal $x$ and $y$ directions. 
This pronounced anisotropy stems from the $d_{x^2-y^2}$-wave AM feature \cite{zhang2025giant}. 
Consistently, the angular dependence of the spin polarization $\mathrm{SP}_\mathbf{nn}$
follows a $\cos(2\varphi)$ dependence (see \textcolor{blue}{Fig.~S5} \cite{supp}), making it difficult to ensure robust 
spin polarization in AM systems.

\begin{figure*}[htbp]
        \centering
        \includegraphics[width=0.9 \linewidth]{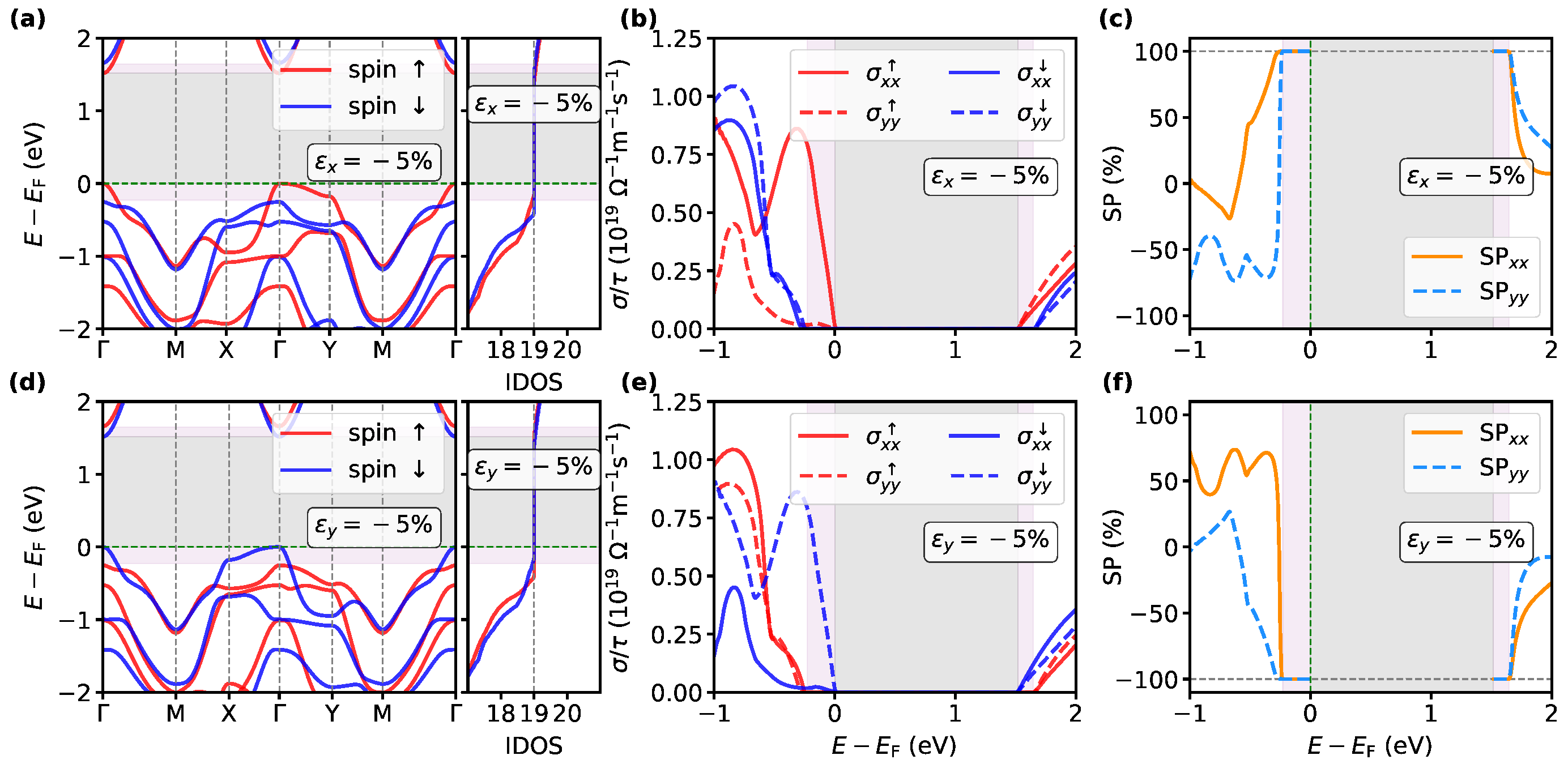}
        \caption{(a) Band structure and IDOS, (b) spin-resolved conductivity as a function of energy, 
	and (c) energy-dependent spin polarization along the $x$ and $y$ directions for Mn$_2$SeO 
        under 5\% compressive strain along the $x$ axis ($\varepsilon_x$ = -5\%). 
        (d)–(f) The corresponding results under 5\% compressive strain along the $y$ axis ($\varepsilon_y$ = -5\%), 
        shown for direct comparison and highlighting the strain-induced spin reversal.
    The light gray and pink shaded areas represent the band gap  
    and spin-splitting gap regions, respectively.	        
        }
        \label{fig3}
\end{figure*}

Based on the model analysis above, uniaxial strain can be introduced 
to lift symmetry constraints and facilitate spin splitting, 
with 5\% compressive strain along $x$ and $y$ directions ($\varepsilon_x$ = -5\% and $\varepsilon_y$ = -5\%)
taken as representative cases.
For $\varepsilon_x$ = -5\% case, ML-Mn$_2$SeO undergoes a magnetic phase transition 
from AM to fFIM, evidenced by IDOS present in Fig.~\ref{fig3}(a). 
The band structure exhibits pronounced and anisotropic spin splitting throughout the BZ.
Near $\Gamma$ point, spin-up valence bands lie above the spin-down ones, whereas in the conduction band region spin-up states are pushed below the spin-down counterparts. 
As a result, spin-splitting gaps of about 0.23 and 0.13~eV are formed, respectively, reflecting 
a half-semiconducting behavior consistent with the model results 
illustrated in Fig.~\ref{fig1}(f) and~\ref{fig1}(h).
This band feature leads to spin-up dominated transport within the spin-splitting gap region along 
both $x$ and $y$ directions, as further confirmed by the calculated spin-resolved 
longitudinal conductivities [see Fig.~\ref{fig3}(b)]. 
We also see that the conductivities become strongly anisotropic, with broken spin equivalence 
between $\sigma_{ii}^{\uparrow}$ and $\sigma_{ii}^{\downarrow}$ ($i=x,y$), signaling 
the lifting of symmetry constraints. Correspondingly, a full spin polarization of +100\% around 
VBM or CBM is
achieved  along both principal directions [see Fig.~\ref{fig3}(c)].

In contrast, as displayed in Fig.~\ref{fig3}(d), applying a compressive strain of 5\% 
along the $y$ direction ($\varepsilon_y$ = -5\% ) yields a similar AM-to-fFIM transition 
but with a reversed spin texture. 
The spin-resolved bands retain similar splitting features, whereas the spin polarization near the $E_\mathrm{F}$
is inverted accompanied by a $C_{4z}$ rotation of the band path 
compared to $\varepsilon_x$ = -5\% configuration shown in Fig.~\ref{fig3}(a),
becoming fully dominated by spin-down states.
This spin inversion is consistently reflected in spin-resolved conductivities [see Fig.~\ref{fig3}(e)], 
where the full conductivity tensor undergoes a simultaneous exchange of 
spin indices and a permutation between $x$ and $y$ components. 
Consequently, the spin polarization $\mathrm{SP}$ reaches -100\% [see Fig.~\ref{fig3}(f)].
These reversals are rooted in the fact that these two configurations 
are connected by the [$C_2 \Vert C_{4z}$] symmetry operation.
It is emphasized that, this uniaxial strain-controlled reversible full 
spin polarization does not rely on the direction of an external electric 
field (see \textcolor{blue}{Fig.~S6} \cite{supp}),
implying a deterministic route to reverse the spin polarization 
via strain engineering.

\begin{figure*}[htbp]
        \centering
        \includegraphics[width=0.9 \linewidth]{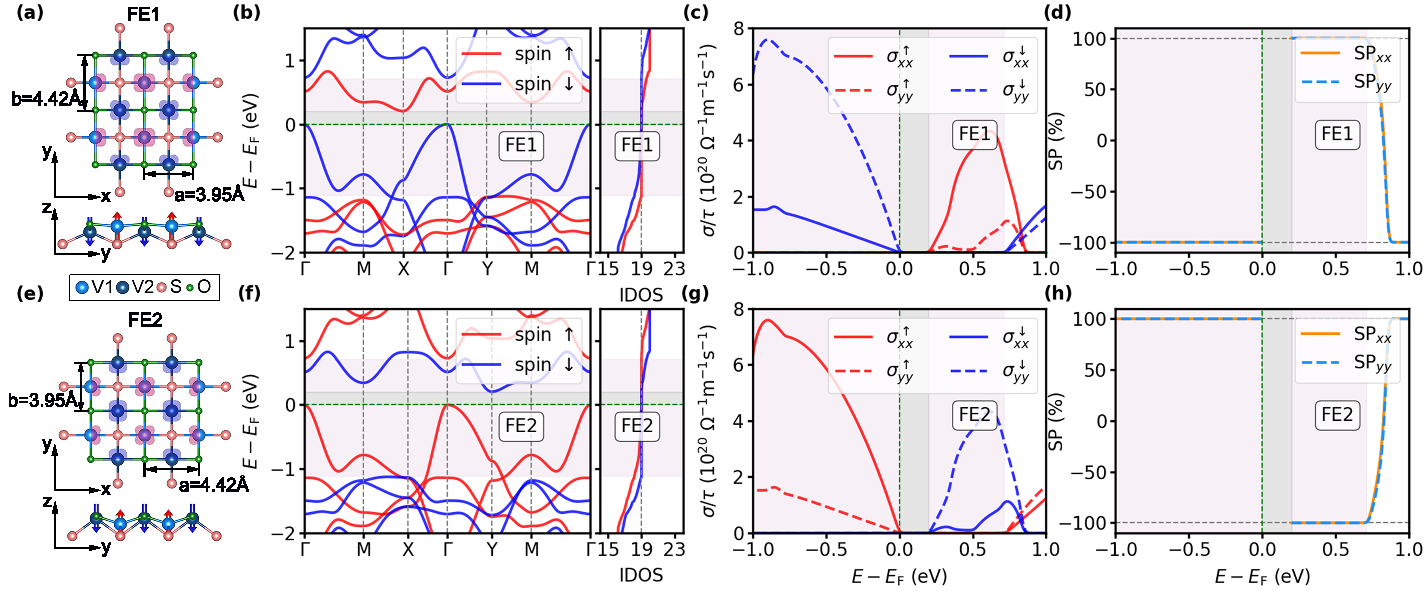}
	\caption{(a) Crystal structure and spin density, (b) spin polarized band structure and IDOS, (c)
	spin-resolved conductivity, and (d) spin polarization SP along the $x$ and $y$ directions 
	for the FE1 phase of V$_2$SO. (e)–(h) The corresponding quantities for the FE2 phase, 
	shown in the same sequence for direct comparison.
    The light gray and pink shaded areas represent the band gap  
    and spin-splitting gap regions, respectively.	
	}
        \label{fig4}
\end{figure*}

\textbf{Ferroelastic fFIM material V$_2$SO}. 
To demonstrate the generality of the proposed mechanism, we further consider the 
monolayer V$_2$SO (ML-V$_2$SO) as the other representative candidate. 
Distinct from ML-Mn$_2$SeO, the ground-state structure of ML-V$_2$SO adopts a rectangular lattice 
belonging to the $Pmm2$ space group (No.~25), instead of the higher-symmetry $P4mm$ phase, 
and is described by the $C_{2v}$ point-group symmetry.
As illustrated in 
Figs.~\ref{fig4}(a) and~\ref{fig4}(e), ML-V$_2$SO hosts two energetically 
degenerate ferroelastic configurations, denoted as FE1 ($a<b$) and FE2 ($a>b$), respectively,
with inequivalent in-plane lattice constants of 3.95~{\AA} and 4.42~{\AA},
as illustrated in Figs.~\ref{fig4}(a) and~\ref{fig4}(e).
Magnetically, two inequivalent V sites (V1 and V2) carry antiparallel magnetic moments forming a Néel order. 
However, unlike AM states, V1 and V2 are not connected by any symmetry operation. Despite the vanishing net 
magnetization, this absence of symmetry relation gives rise to a fFIM state, intrinsically coupled 
to the ferroelastic degree of freedom. 
Notably, FE1 and FE2 configurations in ML-V$_2$SO are related by the [$C_2 \Vert C_{4z}$] symmetry
and can be interconverted via applying the uniaxial strain. The transition is mediated by 
a higher-symmetry paraelastic AM phase with $P4mm$ space-group symmetry \cite{jql2-b4c4,t774-62jc}. 
As a result, the overall switching process follows the theoretical framework established in Fig.~\ref{fig1}(e) and \ref{fig1}(g).

Figure~\ref{fig4}(b) displays the electronic band structure of ML-V$_2$SO in the FE1 configuration. 
It is shown that the valence bands are fully spin-down polarized with a very large spin-splitting gap of $\sim$1.12~eV, 
whereas conduction bands are entirely dominated by spin-up states with a spin-splitting gap of $\sim$0.52~eV. 
Such a band alignment is characteristic of a bipolar magnetic semiconductor \cite{PhysRevB.111.205407,PhysRevB.102.081402}, 
indicating strong spin selectivity. 
These features are consistently reflected in the energy-dependent spin-resolved conductivities 
and spin polarization [see Fig.~\ref{fig4}(c) and~\ref{fig4}(d)]. 
As a consequence, electrostatic gating can be employed 
to tune the carrier type \cite{jiang2018controlling,Liu_2024}, 
enabling fully spin-polarized transport and reversible 
spin polarization switching. 
Moreover, the large spin-splitting gaps provide an extended energy window, significantly 
enhancing the robustness and tunability of spintronic operations.
In contrast, FE2 retains spin-splitting feature but reverses the spin character relative to FE1 case [see Fig.~\ref{fig4}(f)]. Correspondingly, the spin-resolved conductivities exhibit a symmetry-related interchange 
of spin indices and $x/y$ components, leading to a fully spin-polarized state 
with opposite sign [see Figs.~\ref{fig4}(g) and
\ref{fig4}(h)].
These results confirm that ferroelastic switching drives a robust AM-to-fFIM transition, 
enabling both the enhancement and deterministic reversal of spin polarization, 
thereby providing a universal route for strain-controlled spin transport.

\textit{Conclusion.} 
In summary, we establish a symmetry-based mechanism for achieving fully spin-polarized transport 
and its reversible switching in magnetically compensated systems. It is found that
the uniaxial strain can lift the symmetry constraints in AM systems, 
induces an AM-to-fFIM transition, and enables full spin polarization through the emergence 
of large spin-splitting gaps.
Beyond this transition, strain along orthogonal directions produces two symmetry-related and 
energetically equivalent fFIM states with opposite spin polarization, corresponding to ferroelastic phases. 
This reveals that the essential physics is not limited to altermagnets, but can be generalized to 
ferroelastic fFIM systems, where spin reversal is intrinsically tied to structural symmetry.
These predicted behaviors are validated in 2D AM Mn$_2$SeO and ferroelastic V$_2$SO, 
demonstrating perfect spin polarization and its deterministic reversal. 
These results pave a unified route toward symmetry-controlled, volatile or nonvolatile, 
and energy-efficient spintronic functionalities.

\section{acknowledgments}
The work was supported by the National Natural Science
Foundation of China (Grant Nos. 2574065, 12204202, and 11974355);
the Natural Science Foundation of Jiangsu Province (Grant
No. BK20220679).

\bibliography{ref}

\end{document}